\documentclass[lettersize,journal]{IEEEtran}
\usepackage{amsmath,amsfonts}
\usepackage{algorithmic}
\usepackage{algorithm}
\usepackage{array}
\usepackage[caption=false,font=normalsize,labelfont=sf,textfont=sf]{subfig}
\usepackage{textcomp}
\usepackage{stfloats}
\usepackage{url}
\usepackage{verbatim}
\usepackage{graphicx}
\usepackage{cite}
\usepackage{cellspace}
\usepackage{caption}
\usepackage{makecell}
\usepackage{booktabs}
\usepackage[inkscapelatex=false]{svg}
\usepackage{amsmath,amssymb,amsfonts,}
\usepackage{svg} 
\bstctlcite{IEEEexample:BSTcontrol}

\begin{document}

\title{Wireless Environment Information Sensing, Feature, Semantic, and Knowledge: Four Steps Towards 6G AI-Enabled Air Interface}

\author{Jianhua Zhang, Yichen Cai, Li Yu, Zhen Zhang, Yuxiang Zhang, Jialin Wang, Tao Jiang, Liang Xia, Ping Zhang
}



\maketitle

\begin{abstract}
The air interface technology plays a crucial role in optimizing the communication quality for users. To address the challenges brought by the radio channel variations to air interface design, this article proposes a framework of wireless environment information-aided 6G AI-enabled air interface (WEI-6G AI$^{2}$), which actively acquires real-time environment details to facilitate channel fading prediction and communication technology optimization. Specifically, we first outline the role of WEI in supporting the 6G AI$^{2}$ in scenario adaptability, real-time inference, and proactive action. Then, WEI is delineated into four progressive steps: raw sensing data, features obtained by data dimensionality reduction, semantics tailored to tasks, and knowledge that quantifies the environmental impact on the channel. To validate the availability and compare the effect of different types of WEI, a path loss prediction use case is designed. The results demonstrate that leveraging environment knowledge requires only 2.2 ms of model inference time, which can effectively support real-time design for future 6G AI$^{2}$. Additionally, WEI can reduce the pilot overhead by 25\%. Finally, several open issues are pointed out, including multi-modal sensing data synchronization and information extraction method construction.

\end{abstract}


\section{Introduction}

\IEEEPARstart{T}{he} deep integration of artificial intelligence (AI) with the sixth-generation (6G) mobile communication system empowers numerous smart application scenarios. With in a communication systems, air interface refers to the communication interface facilitating data transmitted between the base station (BS) and the terminal. In Release 18, the 3rd Generation Partnership Project (3GPP) delves into AI-enabled air interface (AI$^{2}$) \cite{38843}, which will also be a key focus in future standards studies for radio access networks (RAN).

As shown in Fig. \ref{arc}, data sent by the source is modulated and encoded for transmission through the radio channel, and reaches the sink after demodulation and decoding. In this process, the radio channels dramatic varying in time-, frequency-, and spatial-dimensions determines the system performance. To combat the variable channel fading, the transmitter (Tx) needs to design the signal processing methods based on the current channel quality to improve the link performance. Within this framework, AI-based channel prediction methods can efficiently acquire channel fading status and achieve end-to-end automatic optimization, thereby enhancing transmission efficiency, reducing resource consumption, and minimizing complexity. In \cite{ZZ_AI}, deep learning (DL) models are used to leverage correlations across multiple dimensions of the channel and can reduce the number of pilots transmitted across different antennas, time slots, and frequency bands. Yet, such data-driven algorithms are trained offline with scenario-specific data, posing challenges when the scenario changes significantly or is applied to a new scenario. 

So, what factors in different scenarios will affect the channel? Electromagnetic (EM) waves emitted from the Tx encounter buildings, vegetation, and other scatterers, leading to multipath transmission. The interaction of these EM waves traveling along different paths causes channel fading. The further the scatterer is, the longer the multipath propagation delay is. The larger the size of the scatterers is, the wider its shadowing area is. Additionally, a denser scatterer layout leads to an increased number of multipaths, as well as an increase in delay spread and angular spread.
The quality of the radio channel is ultimately determined by how EM waves interact with objects within it \cite{cost}.

Therefore, we define wireless environment information (WEI) as encompassing all environmental objects that determine channel variations, such as the position, size, and layout of the scatterers. In the past, communication systems were unable to obtain real-time WEI. Fortunately, the rapid development of 6G sensing technology provides a solution to this problem. Multi-modal sensing technology is capable of acquiring RGB images \cite{robert}, depth images \cite{PEACH}, point clould \cite{point-clould}, and other information that contains WEI. This inspires us to take a new approach, acquiring the channel fading status of specific sites from WEI in real-time, referred to as environment-channel mapping. When applied to the 6G AI$^{2}$, the spatial distribution of scatterers can be recognized by sensing, thus allowing AI models to learn the fundamental mapping from the environment to the channel. The incorporation of WEI helps AI models maintain stable performance when generalized to different scenarios.

During the process of environment-channel mapping, reducing the volume of raw sensing data and finding a more concise representation of WEI can reduce storage space and computational costs. In existing research, low-dimensional WEI representations include environment features such as scatterer size \cite{SYT_feature}, environment graph \cite{SYT_GNN}, and environment semantics such as scatterer layout information \cite{SYT_sem}, effective scatterers heatmap \cite{GFF_sem}, and terminal mask \cite{ahmed_sem}. Besides, there is environment knowledge that directly characterizes wireless channel attributes, such as radio environment knowledge (REK) \cite{REKP_pl}, and channel knowledge map \cite{ZY_map}.

However, the different kinds of WEI in these studies are discrete and unrelated. To clarify the distinctions among the diverse WEI concepts and expound on their respective roles in the AI$^{2}$, this article propose a framework of WEI-aided 6G AI$^{2}$ (WEI-6G AI$^{2}$). Four progressive steps for obtaining WEI are established, in which the accurate extraction of former step information can facilitate the acquisition of latter step information. After the WEI has evolved to the final step, it can provide AI-based prediction methods with channel-related prior knowledge in the most effective way, helping to obtain accurate channel fading status or to optimize transmission techniques with low overhead in a varying environment, thus truly moving 6G AI$^{2}$.

The organization of this article is as follows. Section II gives a framework of WEI-6G AI$^{2}$. Section III establishes four steps for obtaining WEI and clarifies their connections and differences. Section IV gives a case study of using WEI for path loss (PL) prediction and channel state information (CSI) prediction. Section V discusses several challenges encountered in the application process. Section VI gives the conclusion and outlines future research directions.

\section{Framework}

\begin{figure*}[t!]
    \centering
    \includegraphics[width=0.95\textwidth]{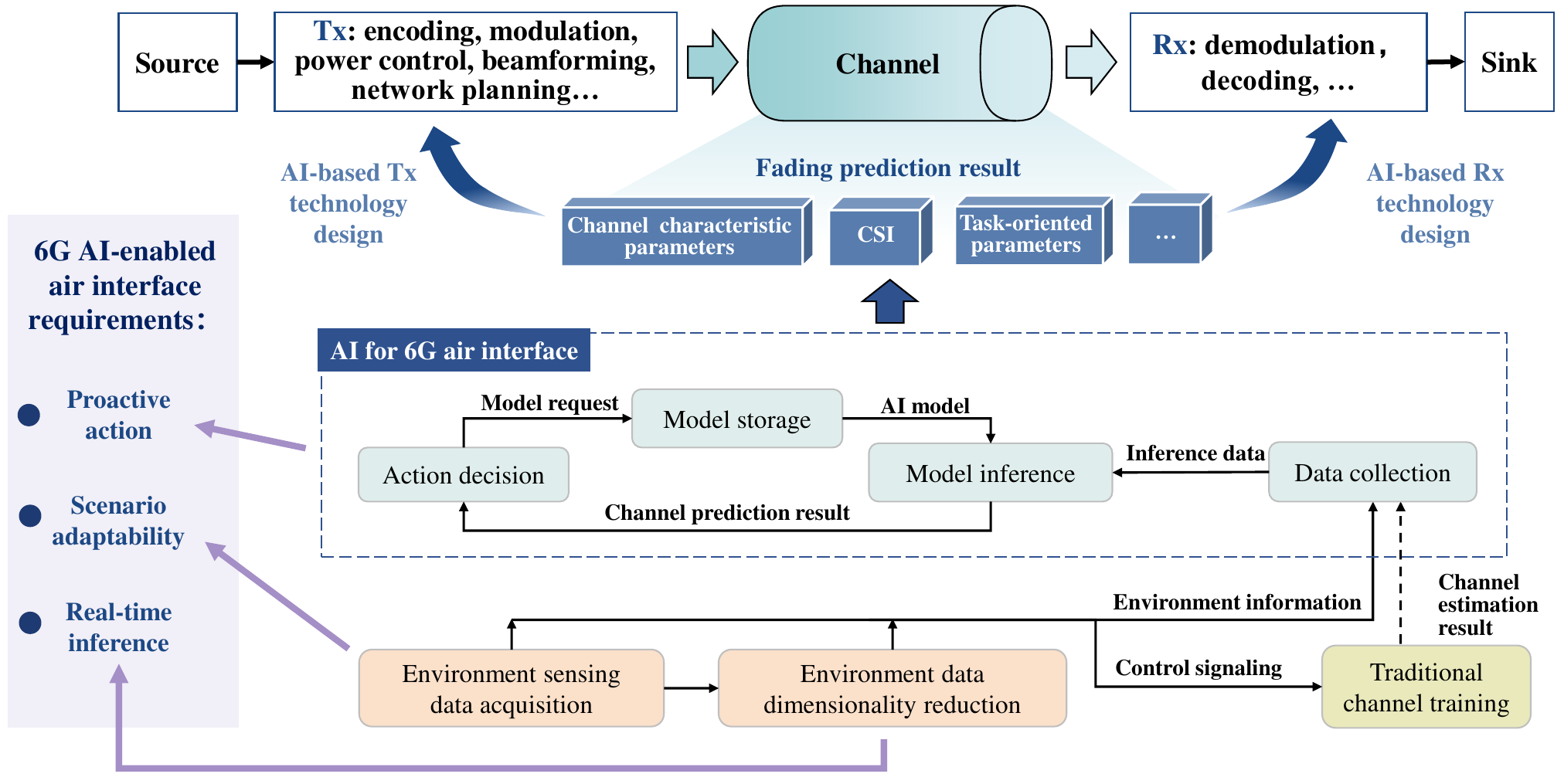}
    \captionsetup{justification=centering}
    \caption{WEI-6G AI$^{2}$ framework for 6G network}
    \label{arc}
\end{figure*}

The goal of the AI$^{2}$ is to provide communication data in the most efficient way possible within the constraints of the available hardware and the radio environment\cite{ai-ai}. Therefore, we believe that the following requirements should be satisfied:

\begin{itemize}
    \item \textbf{Scenario adaptability:} AI$^{2}$ is expected to be utilized in optimizing communication schemes across various scenarios such as indoor office, indoor factory, urban macro, etc. Consequently, AI models applied to specific tasks need to be scenario-adaptive without consuming time to retrain during deployment.
    \item \textbf{Real-time inference:} Traditional methods apply offline statistical channel models for system design, and the actual performance of the algorithms is far from the design expectations. To address this limitation, AI$^{2}$ needs to reason in real-time inference within a frame (10 ms) and obtain the channel fading status quickly online to achieve performance improvement.

    \item \textbf{Proactive action:} Unlike traditional communication systems passively adapt to the environment, the AI$^{2}$ actively and automatically designs signal processing schemes through the acquired channel fading status to optimize the effectiveness and reliability of the data transmission under the constraints of the current environment.
    
\end{itemize}

In order to realize these requirements, we propose the framework of WEI-6G AI$^{2}$, which consists of four parts: environment sensing data acquisition, environment data dimensionality reduction, traditional channel training, and AI for air interface, as shown in Fig. \ref{arc}.

\subsection{Environment Sensing Data Acquisition}
Sensing data is used to enhance the scenario adaptability of the AI model. When the Tx position is fixed, there is a correlation in the channel fading status between receivers (Rxs) that are close to each other, due to the fact that their channels experience similar physical environments. When WEI is not considered, AI in the air interface learns the environment-specific mapping function between channels, making it challenging for this mapping relationship to be equally applicable to new scenarios. In the proposed WEI-6G AI$^{2}$ that takes sensing data into account, instead of needing to implicitly learn about the environment through the channel-channel mapping relationship, deep learning (DL) based AI algorithms can learn the environment-channel mapping relationship by directly seeing the current scatterer state around Tx and Rx. In this way, when the environment changes, a new channel fading status can be obtained using the new sensing result, which enhances the portability of the AI model.

The system employs different types of sensing devices to provide multi-modal sensing data, which is then used to reconstruct a three-dimensional (3D) environment model. Since static objects such as buildings, plants, and roads do not change over time, their states do not need to be tracked in real time. Therefore, The static map can be captured in the early stage by unmanned aerial vehicles (UAVs) and stored in the BS. In practical deployments, dynamic scatterers such as pedestrians and vehicles are extracted in real-time with target detection and video tracking methods. Finally, the dynamic scatterers are added to the static map to form a complete 3D environment model.

\begin{figure*}[t!]
    \centering
    \includegraphics[width=0.95\textwidth]{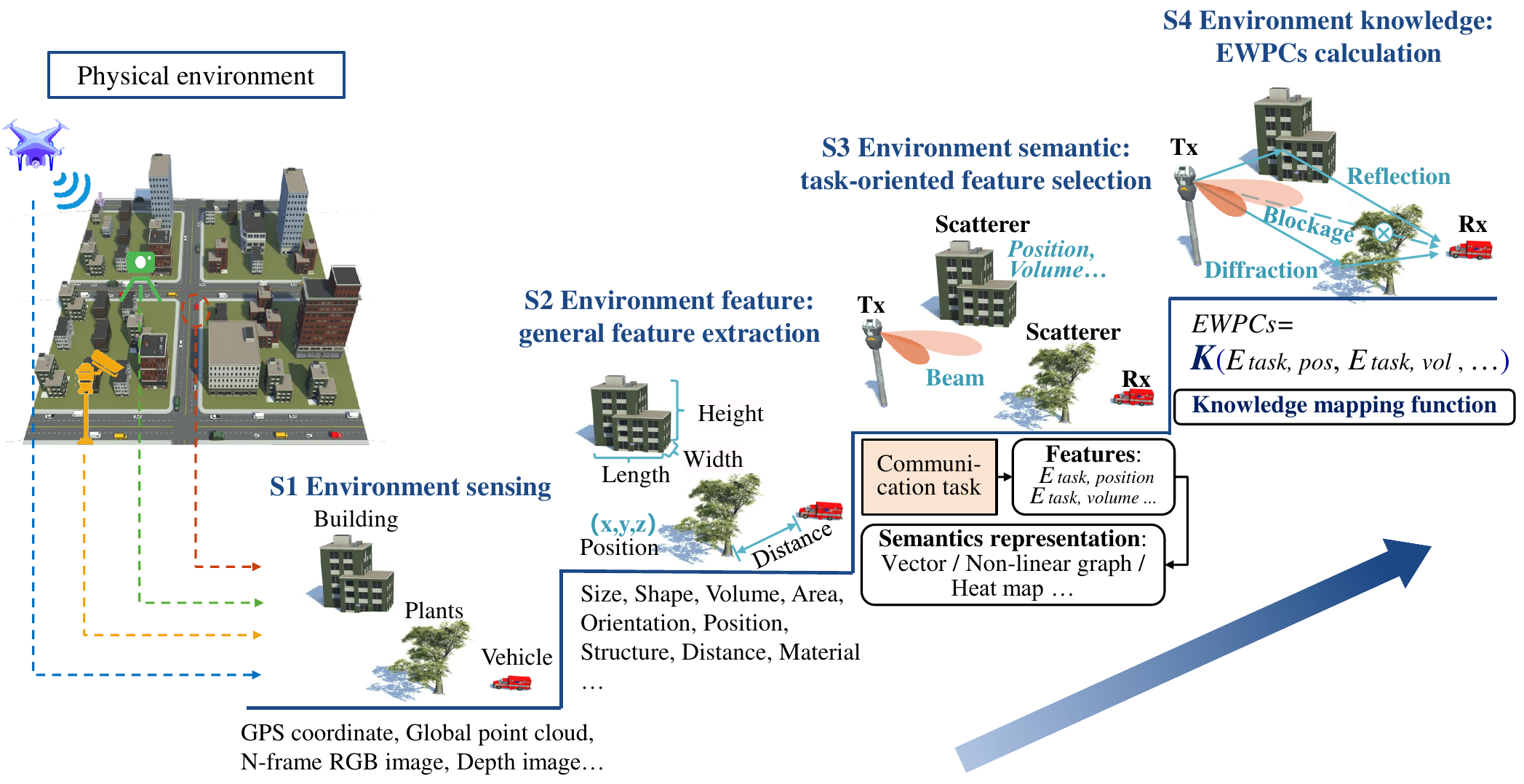}
    \captionsetup{justification=centering}
    \caption{Four progressive steps of WEI towards 6G AI$^{2}$}
    \label{4level}
\end{figure*}

\subsection{Environment Data Dimensionality Reduction}

Data dimensionality reduction is used to support real-time inference of AI$^{2}$. Sensing data provides a more accurate environment awareness for wireless communication systems, but its data volume is enormous. In \cite{point-clould}, a 7.20 × 6.07 × 2.56 $m^{3}$ ultrasonic inspection room is measured using a laser scanner with a point cloud model containing a total of 68,848 points (equivalent to an average of 614 points per $m^{3}$). In scenarios where outdoor users and BS are tens of meters away, the amount of data required to preserve the point cloud will become prohibitive. As the dimensionality of the sensing data increases, the computational complexity of the  DL network increases exponentially, and the inference process becomes extremely time-consuming.

In the proposed WEI-6G AI$^{2}$, data dimensionality reduction is achieved by information extraction from raw sensing data, and can reduce the inference time of AI$^{2}$. Specifically, we first conduct an in-depth analysis of the environmental factors affecting channel fading, including but not limited to multipath effects, interference conditions, or the locations and materials of effective scatterers. Afterward, different methods for extracting this information from the reconstructed 3D environment model are built, and constructed as a methods pool. 

Once constructed, the information extraction methods remain unchanged throughout the entire deployment process. However, the extracted WEI will be regularly updated at a certain frequency. This module will output control signaling or low-dimensional WEI, to subsequently change the pilot sending scheme, or feed into machine learning (ML), DL, reinforcement learning, or other AI prediction models to achieve the communication task.

\subsection{Traditional Channel Training}
In the proposed WEI-6G AI$^{2}$, WEI is not only used for channel fading prediction, but also has an impact on the channel training process. When the required parameters are closely related to the environment, we can skip the process such as downlink pilot transmission, feedback, or beam scanning, and directly predict the needed parameters using WEI. 

When the required parameters are difficult to find an intuitive connection to the environment, The AI model in the WEI-6G AI$^{2}$ framework does not infer channel parameters directly from WEI, and we still need to transmit some pilots for channel estimation. In this process, the pilot is used to provide some of the channel information while WEI is used to reduce the uncertainty of the remaining channel information. Regardless of the task requirements, WEI can help to reduce communication resource overhead.

\subsection{AI For 6G Air Interface}

When applied to different tasks, the channel fading status is manifested in different parameters, including channel large-scale parameters (LSP) such as PL, shadow fading, delay spread, angle spread, channel small-scale parameters (SSP) such as multipath power, delay, angle, or task-oriented parameters such as the optimal beam index. In the proposed WEI-6G AI$^{2}$, DL and ML models establish a mapping relationship from WEI and fewer channel estimation results to multiple channel parameters, which enables 6G networks to proactively adapt the communication techniques through the environment.

The specific AI model is selected by the action design module according to the communication task, and data collection is performed to support model training and inference. Finally, the decision design module completes system design optimization based on the channel fading prediction results. AI can be used to enhance the performance of individual modules when the original system architecture is not changed and the encoding, modulation, and other modules are designed and optimized separately. In addition to this, AI can also be used for end-to-end optimization. For example, the author of \cite{end_end_AI} uses a deep convolutional neural network (CNN) to replace the entire received signal processing consisting of channel estimation, equalization, and soft demapping.

In conclusion, WEI-6G AI$^{2}$ enables the real-time and proactive optimization of MIMO transmission, coding and modulation, wireless energy transfer, and other technologies according to the channel fading status while reducing the dependence on pilot transmission. It also supports personalized services for different users, allowing the network to dynamically allocate resources and optimize links based on user requirements, thus enhancing overall communication quality and user experience.

\section{evolution of WEI}
In the proposed WEI-6G AI$^{2}$, The WEI output from the environment sensing data acquisition module and the environment data dimensionality reduction module has a variety of representations. As shown in Fig. \ref{4level}, the acquisition of WEI can be divided into four steps. The differences between WEIs obtained at each step are summarised in Tab. \ref{tab1}. 

\begin{table*}
\begin{center}
\caption{Characteristics of different WEIs and their supporting roles in 6G AI$^{2}$}
\label{tab1}
\renewcommand{\arraystretch}{1.5}
\setlength\cellspacetoplimit{4pt}
\setlength\cellspacebottomlimit{4pt}
\begin{tabular}{| Sc | Sc | Sc | Sc | Sc |}
\hline
\bfseries Attributes & \bfseries $S_1$ Sensing data& \bfseries $S_2$ Feature & \bfseries $S_3$ Semantic & \bfseries $S_4$ Knowledge\\
\hline
Dimensionality reduction& Exclude & Include& Include& Include\\
\hline
Deconstruction method& \makecell[c]{Directly use \\neural network} & \makecell[c]{Global-oriented\\Local-oriented} & Task-oriented & Task-oriented\\ 
\hline
\makecell[c]{Type of the deconstructed\\ information}& Inexplicable & \makecell[c]{Environmental \\characteristics} & \makecell[c]{Environmental \\characteristics} & \makecell[c]{Electromagnetic wave\\ propagation characteristics} \\

\hline
Scenario adaptability & Moderate & Moderate & Moderate & Significant \\
\hline
Real-time inference & Minor & Moderate & Moderate & Significant \\
\hline
Proactive action & Moderate & Moderate & Significant & Significant \\

\hline 
\end{tabular}
\end{center}
\end{table*}

\subsection{Step1: Environment Sensing}

Step-1 ($S_1$) directly utilizes the sensing data obtained from various sensors, including user location coordinates provided by GPS, depth images from depth cameras, red-green-blue (RGB) images from cameras, and point clouds from LiDAR.

After simple preprocessing such as data cleaning and normalization, $S_1$ sensing data is directly put into a suitable neural network (NN) for task-relevant feature vector extraction and channel parameter prediction. CNN-based networks are suitable for processing image data, while point cloud data can be processed using PointNet. The $S_1$ sensing data is easily accessible and their application methods are conceptually simple, thus they can be useful in most prediction tasks such as LSP, SSP, and task-oriented prediction.

The limitations of $S_1$ sensing data are also very obvious. Not all objects in the physical environment affect wireless propagation, leading to a significant amount of redundancy in the sensing data. Directly inputting a large amount of $S_1$ sensing data into NNs will increase the computation time and potentially cause issues such as overfitting, thereby impacting the accuracy of channel fading prediction. Furthermore, $S_1$ sensing data relies solely on NNs for the extraction of task-relevant feature vectors. The intricate workings of NNs make it challenging to clearly explain the correspondence between the extracted features and the actual environmental conditions.

\subsection{Step2: Environment Feature}

In order to reduce the computation time for real-time inference, Step-2 ($S_2$) performs a generic feature extraction of the sensing data to compress the data volume. 

$S_2$ features firstly subdivide the sensing information to distinguish different scatterers in the environment and identify their center coordinates, vertex coordinates, etc. For example, we extracted the scatterer's position and size as environment features for PL prediction in \cite{SYT_feature}. In addition to such basic properties of scatterers, the layout feature is important as well. In \cite{SYT_GNN}, we used non-linear maps to preserve the complete structure of the environment to fully account for the global layout of the objects within the communication environment. 

To sum up, $S_2$ feature is defined as global environment features or local environment details extracted within the scene, such as distance features, geometric features, structural features, or blockage of Line-of-Sight (LoS). The accurate extraction of these basic environment features will provide the basis for the generation of latter step information. However, $S_2$ feature only gives a generalized approach to feature extraction under one line of thought, which is not specific to a particular task. This makes the channel-related features may be redundant or missing.

\begin{figure*}[t!]
    \centering
    \includegraphics[width=0.95\textwidth]{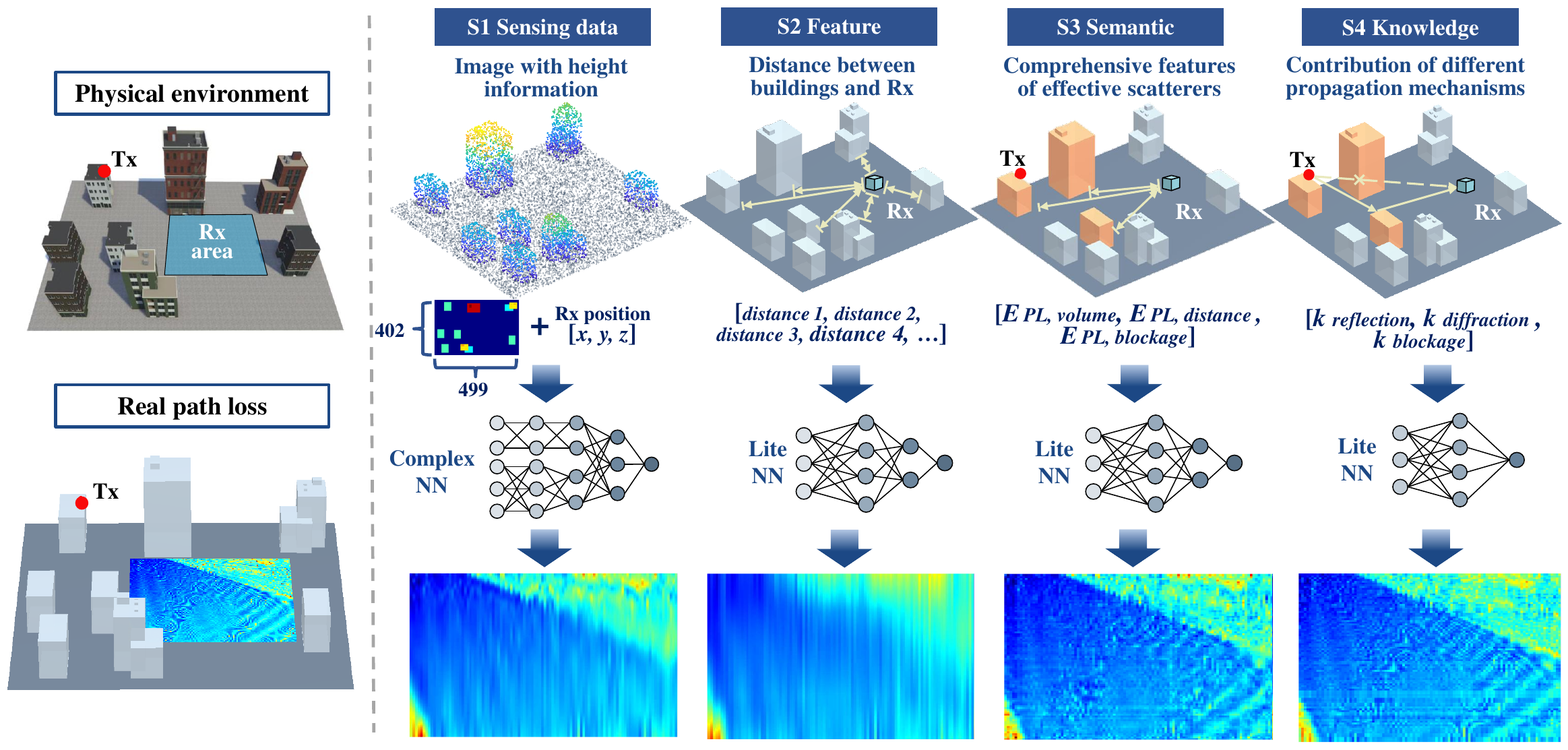}
    \captionsetup{justification=centering}
    \caption{Large-scale channel fading prediction result based on 4-step WEI}
    \label{pl}
\end{figure*}

\subsection{Step3: Environment Semantic}
Step-3 ($S_3$) establishes associations between $S_2$ features and channel prediction tasks, aiming to provide a more nuanced and task-oriented understanding of the environment, optimizing the use of data for precise and efficient action.

Recognizing that different environment features have varied importance for different application tasks, the $S_3$ semantics is constructed by choosing relevant environment features\cite{SYT_sem}. We denoted the required $S_2$ features by $\emph{E}_{task, feature}$, and constructed the task-oriented environment semantics as the combination of $\emph{E}_{task, feature}$ in various forms, such as vectors, nonlinear graphs, etc. For example, since LoS blockage may cause communication interruptions due to the large received power degradation, $S_3$ semantics can be constructed by combining LoS blockage features with scatterer layout features when performing beam prediction.

However, the $S_3$ semantics remains at the stage of describing the characteristics of the environment. This means that the DL network needs to implicitly learn the electromagnetic wave propagation characteristics (EWPCs) from the environment characteristics and map them to the channel fading status or task parameters. Since the AI model is black-box and cannot clearly explain whether the mapping relationships it learns are consistent with the physical laws, this uncontrollable prediction makes the trained network prone to errors when applied to new scenarios.

\subsection{Step4: Environment Knowledge}
In order to enhance the scenario generalization of AI, step-4 ($S_4$) further computes based on features and semantics to describe the characteristics of electromagnetic wave propagation (EWP) from the perspective of the environment. Specifically, we first extract a portion of the EWPCs from the environment data in an interpretable way, and then combine them with the AI to achieve the communication task. The mapping relationship that AI needs to learn in this process is relatively simple and easier to generalize to other scenarios.

We define REK as a quantitative analysis of the relationship between effective environment information and EWPC, and use $K(\cdot)$ to represent the process of REK expression \cite{REKP_pl}. The accurate extraction of $S_2$ feature and $S_3$ semantics such as the basic information of scatterers, Tx and Rx positions, and effective scatterer index in the physical environment allows us to compute the $S_4$ knowledge by theoretical derivation in terms of EWP principles, fast fading model, stochastic geometry, and so on.

WEI obtained in the latter step can better meet the  WEI-6G AI$^{2}$ requirements, and applying $S_4$ knowledge can bring the following advantages:
\begin{itemize}
\item \textbf{Reduces channel uncertainty:}
$S_4$ Knowledge performs data dimensionality reduction through interpretable mathematical methods, reducing the uncertainty of the EWP process. This enables the $S_4$ knowledge to extract a greater amount of channel fading-related information from the same sensing data, assisting the system to make more accurate actions. 

\item \textbf{Provide efficient environment representation:} The evolution of WEI from $S_1$ to $S_4$ accomplishes the deconstruction and dimensionality reduction of the sensing data, allowing $S_4$ knowledge to provide concise data for interactions between different parts of the system, thereby improving the efficiency of information transmission. 

\item \textbf{Achieve lite network:} The information extracted by REK is directly related to the channel fading status. Therefore, establishing the mapping of $S_4$ knowledge with a communication task does not require a complex DL model with powerful mapping capabilities. This helps to achieve a lite network, which reduces the inference latency of the AI prediction model.
\end{itemize}

\begin{table*}
\begin{center}
\caption{Comparative of PL prediction using different WEI}
\label{tab2}
\renewcommand{\arraystretch}{1.5}
\setlength\cellspacetoplimit{4pt}
\setlength\cellspacebottomlimit{4pt}
\begin{tabular}{| Sc | Sc | Sc | Sc | Sc |}
\hline
\bfseries Attributes & \bfseries $S_1$ Sensing data& \bfseries $S_2$ Feature & \bfseries $S_3$ Semantic & \bfseries $S_4$ Knowledge\\

\hline
Specific information & \makecell[c]{Top view hight image \\ \& Rx GPS coordinate} & \makecell[c]{Distance between each\\ building and Rx}& \makecell[c]{Volume, distance, and\\ quantified  LoS blockage \\of effective scatterers} & \makecell[c]{Quantified reflection,\\ diffraction, and \\blockage contributions}\\
\hline
\makecell[c]{Quantities of WEI data\\ per Rx} & \makecell[c]{499 × 402 (number of pixels)\\ + 3 (xyz-coordinates)} & 10 (number of buildings) & 3 & 3\\ 
\hline
Layers of the NN & \makecell[c]{CNN × 5 \\ Liner × 2} & \makecell[c]{CNN × 3 \\ Liner × 1} & \makecell[c]{CNN × 3 \\ Liner × 1} & \makecell[c]{CNN × 2 \\ Liner × 1} \\

\hline
inference time (s) & 0.4308	& 0.0064 & 0.0042 &	0.0023 \\

\hline
MSE & 9.87 & 11.49 & 7.54 & 5.65 \\

\hline 
\end{tabular}
\end{center}
\end{table*}

\section{simulation}
In this section, both large-scale and small-scale channel fading prediction examples of WEI-6G AI$^{2}$ are performed in simulation, to verify the application effects of different WEI.

\subsection{Dataset Construction and Model Training}
As shown in the left side of Fig. \ref{pl}, an outdoor scenario is considered for a communication system that contains a BS at the top of the building, a mobile user moving in the square area, and some sensing devices. Afterward, we extracted the 4-step WEI from the scenario, with specific information types and data quantities summarized in Tab. \ref{tab2}. Subsequently, we constructed a DL model to map the WEI to the large-scale channel fading PL. For the $S_1$ sensing data, we employed a 2-layer CNN and 1 linear layer to extract the task-relevant feature vector from the images. These vectors were then combined with GPS coordinates and utilized for PL prediction using a 3-layer CNN and 1 linear layer. For the $S_2$ feature and $S_3$ semantic, the NN used for feature vector extraction is removed, retaining only the 3-layer CNN and 1 linear layer for PL prediction. Regarding $S_4$ knowledge, we further reduce the use of 1 CNN layer.

\subsection{Performance Evaluation}

The large-scale PL prediction results for different WEIs are shown in Fig. \ref{pl}, with specific model inference time and prediction mean square errors (MSE) summarized in Tab. \ref{tab2}. It can be observed that the prediction results for $S_1$ sensing data and $S_2$ feature can roughly reflect the variation trends of PL in LoS and Non-Line-of-Sight (NLoS) areas, but cannot reflect the detailed changes. In comparison between these two approaches, the $S_2$ feature significantly reduced 99\% of the data quantity and model inference time. However, due to the insufficient correlation between the extracted distance features and the PL prediction task, this lead to a 16\% performance loss. As it is oriented towards identifying effective scatterers for communication tasks and extracting more comprehensive information, the prediction results for $S_3$ semantic and $S_4$ knowledge are noticeably more detailed. Compared with $S_3$ semantics, $S_4$ knowledge can be mapped to PL through a simpler model, saving 45\% of the prediction time while improving the prediction performance by 25\%.

The simulation results prove that $S_4$ knowledge can reduce the network complexity and improve real-time interface while obtaining accurate channel fading status. However, $S_4$ knowledge needs to construct specific knowledge representations for different tasks, requiring a fine-grained design process based on task requirements. Therefore, when applied to tasks that do not require detailed channel information, such as blockage prediction or beam prediction, directly choosing $S_1$ sensing data or $S_2$ feature to complete the communication task is also a simple and effective way. 

\begin{figure}[t!]
    \centering
    \includegraphics[width=0.43\textwidth]{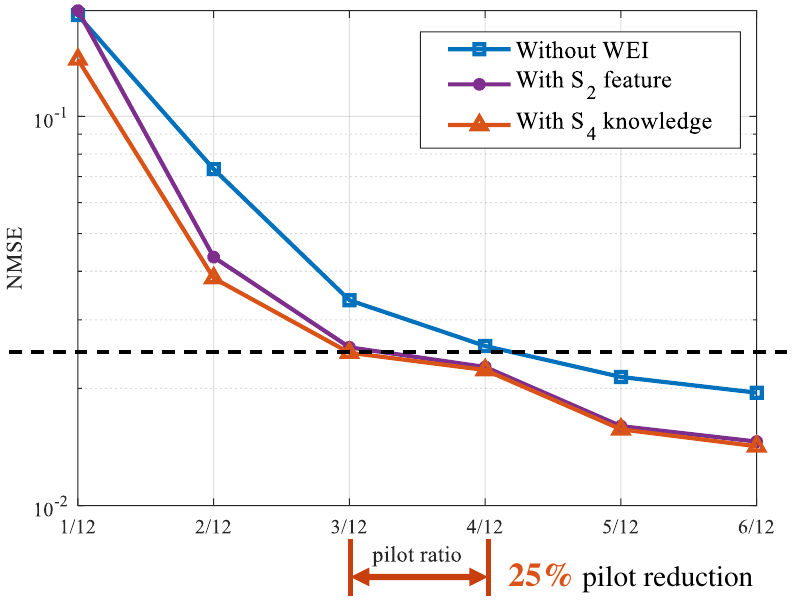}
    \captionsetup{justification=centering}
    \caption{Small-scale channel fading prediction result}
    \label{nmse}
\end{figure}

In addition, we further validate the performance of WEI for small-scale channel fading prediction, using WEI and partial channel estimation results to predict the complete CSI. In Fig. \ref{nmse}, the black line indicates the minimum pilot ratio required by the different algorithms to reach the same performance. It can be seen that WEI can reduce the pilot ratio by 25\%, and that $S_4$ knowledge outperforms the $S_2$ feature at low pilot ratios.

\section{Challenges and future opportunity}
Current results show that using WEI in the 6G AI$^{2}$ is highly advantageous. 
However, there are still some challenges in different parts of WEI-6G AI$^{2}$ that deserve further research.

\textbf{Multi-modal sensing data synchronization:}
The deployed DL models need data collection for model inference. These required sensing data are sourced from various disparate sensing devices, resulting in different data collection delays. This may lead to multi-modal data mismatches, causing lower accuracy in the generated WEI. In the future, there is a need to optimize the time synchronization mechanism to ensure minimal clock deviations between all devices, accurately merging data from different sources.

\textbf{Knowledge construction for various parameter prediction:} 
$S_3$ semantics and $S_4$ knowledge can better support real-time channel fading prediction and accurate action decisions. However, the research on this kind of WEI is still in the early stage, and the constructed semantics and knowledge are only oriented toward tasks such as beam prediction, blockage prediction, and PL prediction, which can intuitively find the connection with the environment. How to construct REK for other prediction tasks will be the focus of future research.

\textbf{Integration with the existing network architecture:}
AI-based algorithms have shown promising performance in various communication tasks such as channel estimation and beamforming. However, the operational mechanisms of current AI$^{2}$ algorithms within the system are not yet clear. Defining the interface between AI, WEI, and the protocol stack, as well as continuously optimizing and updating the deployed AI models, could be a lengthy and complex process.

\section{conclusion}

In order to improve the environment adaptability of AI, this article propose the framework of WEI-6G AI$^{2}$, which includes environment sensing data acquisition, environment data dimensionality reduction, traditional channel training, and AI for the 6G air interface. Subsequently, a 4-step process for obtaining WEI from the environment is summarised, where $S_4$ knowledge has the advantages of reducing channel uncertainty, providing efficient environment representation, and achieving lite network. This can help WEI-6G AI$^{2}$ realize the need for adaptation to the environment, real-time interface, and proactive action. Then, the PL prediction and CSI prediction performance of different WEI is experimentally verified. $S_4$ knowledge is able to achieve the shortest model inference time with the highest prediction accuracy while deducing pilot overhead. Finally, certain challenges and future opportunities in the construction and application of WEI are discussed.

\section{acknowledgement}

This work is supported by the National Key R\&D Program of China (Grant No. 2023YFB2904805), the National Natural Science Foundation of China (No. 62401084), and BUPT-CMCC Joint Innovation Center.

\bibliographystyle{IEEEtran}
\bibliography{ref}

\begin{thebibliography}{10}
\providecommand{\url}[1]{#1}
\csname url@samestyle\endcsname
\providecommand{\newblock}{\relax}
\providecommand{\bibinfo}[2]{#2}
\providecommand{\BIBentrySTDinterwordspacing}{\spaceskip=0pt\relax}
\providecommand{\BIBentryALTinterwordstretchfactor}{4}
\providecommand{\BIBentryALTinterwordspacing}{\spaceskip=\fontdimen2\font plus
\BIBentryALTinterwordstretchfactor\fontdimen3\font minus \fontdimen4\font\relax}
\providecommand{\BIBforeignlanguage}[2]{{%
\expandafter\ifx\csname l@#1\endcsname\relax
\typeout{** WARNING: IEEEtran.bst: No hyphenation pattern has been}%
\typeout{** loaded for the language `#1'. Using the pattern for}%
\typeout{** the default language instead.}%
\else
\language=\csname l@#1\endcsname
\fi
#2}}
\providecommand{\BIBdecl}{\relax}
\BIBdecl

\bibitem{38843}
{3GPP}, ``{Study on Artificial Intelligence (AI)/Machine Learning (ML) for NR air interface},'' \emph{3GPP TR 38.843 V0.1.0.}, Jun. 2023.

\bibitem{ZZ_AI}
Z.~Zhang, J.~Zhang, Y.~Zhang \emph{et~al.}, ``{AI-Based Time-, Frequency-, and Space-Domain Channel Extrapolation for 6G: Opportunities and Challenges},'' \emph{IEEE Vehicular Technology Magazine}, vol.~18, no.~1, pp. 29--39, Mar. 2023.

\bibitem{cost}
M.~Boban and V.~Degli-Esposti, ``{White Paper on Radio Channel Modeling and Prediction to Support Future Environment-aware Wireless Communication Systems},'' \emph{arXiv preprint arXiv:2309.17088}, 2023.

\bibitem{robert}
K.~Patel and R.~W. Heath, ``{Harnessing Multimodal Sensing for Multi-user Beamforming in mmWave Systems},'' \emph{arXiv preprint arXiv:2406.05300}, 2024.

\bibitem{PEACH}
S.~Ayvasik, F.~Mehmeti, E.~Babaians \emph{et~al.}, ``{PEACH: Proactive and Environment-Aware Channel State Information Prediction with Depth Images},'' \emph{Association for Computing Machinery}, vol.~7, no.~1, Mar. 2023.

\bibitem{point-clould}
J.~Järveläinen, K.~Haneda, M.~Kyrö \emph{et~al.}, ``{60 GHz Radio Wave Propagation Prediction In a Hospital Environment Using an Accurate Room Structural Model},'' \emph{2012 Loughborough Antennas \& Propagation Conference (LAPC)}, pp. 1--4, Nov. 2012.

\bibitem{SYT_feature}
Y.~Sun, J.~Zhang, Y.~Zhang \emph{et~al.}, ``{Environment Features-Based Model for Path Loss Prediction},'' \emph{IEEE Wireless Communications Letters}, vol.~11, no.~9, pp. 2010--2014, Sept. 2022.

\bibitem{SYT_GNN}
Y.~Sun, J.~Zhang, Y.~\vspace{0mm}Zhang \emph{et~al.}, ``{Environment Information-Based Channel Prediction Method Assisted by Graph Neural Network},'' \emph{China Communications}, vol.~19, no.~11, pp. 1--15, Nov. 2022.

\bibitem{SYT_sem}
Y.~Sun, J.~Zhang, L.~Yu \emph{et~al.}, ``{How to Define the Propagation Environment Semantics and Its Application in Scatterer-Based Beam Prediction},'' \emph{IEEE Wireless Communications Letters}, vol.~12, no.~4, pp. 649--653, Apr. 2023.

\bibitem{GFF_sem}
F.~Wen, W.~Xu, F.~Gao \emph{et~al.}, ``{Vision Aided Environment Semantics Extraction and Its Application in mmWave Beam Selection},'' \emph{IEEE Communications Letters}, vol.~27, no.~7, pp. 1894--1898, Jul. 2023.

\bibitem{ahmed_sem}
S.~Imran, G.~Charan, and A.~Alkhateeb, ``{Environment Semantic Aided Communication: A Real World Demonstration for Beam Prediction},'' \emph{2023 IEEE International Conference on Communications Workshops (ICC Workshops)}, pp. 48--53, May 2023.

\bibitem{REKP_pl}
J.~Wang, J.~Zhang, Y.~Sun \emph{et~al.}, ``{Electromagnetic Wave Property Inspired Radio Environment Knowledge Construction and AI-based Verification for 6G Digital Twin Channel},'' \emph{arXiv preprint arXiv:2406.00690}, 2024.

\bibitem{ZY_map}
Y.~Zeng, J.~Chen, J.~Xu \emph{et~al.}, ``{A Tutorial On Environment-Aware Communications via Channel Knowledge Map for 6G},'' \emph{IEEE Communications Surveys \& Tutorials}, vol.~26, no.~3, pp. 1478--1519, Feb. 2024.

\bibitem{ai-ai}
J.~Hoydis, F.~A. Aoudia, A.~Valcarce \emph{et~al.}, ``{Toward a 6G AI-Native Air Interface},'' \emph{IEEE Communications Magazine}, vol.~59, no.~5, pp. 76--81, May 2021.

\bibitem{end_end_AI}
M.~Honkala, D.~Korpi, and J.~M.~J. Huttunen, ``{DeepRx: Fully Convolutional Deep Learning Receiver},'' \emph{IEEE Transactions on Wireless Communications}, vol.~20, no.~6, pp. 3925--3940, Jun. 2021.

\end{thebibliography}

\vspace{11pt}


\vspace{-1cm}
\begin{IEEEbiographynophoto}{JIANHUA ZHANG (jhzhang@bupt.edu.cn)}is currently a professor with Beijing University of Posts and Telecommunications.
\end{IEEEbiographynophoto}
\vspace{-1cm}

\begin{IEEEbiographynophoto}{YICHEN CAI caiyichen@bupt.edu.cn)}is currently pursuing the
master's degree with Beijing University of Posts and Telecommunications.
\end{IEEEbiographynophoto}
\vspace{-1cm}

\begin{IEEEbiographynophoto}{LI YU (li.yu@bupt.edu.cn)}is currently a postdoctoral research fellow with Beijing University of Posts and Telecommunications.
\end{IEEEbiographynophoto}  
\vspace{-1cm}

\begin{IEEEbiographynophoto}{ZHEN ZHANG (zhenzhang@imu.edu.cn)}is currently a research fellow with Inner Mongolia University.
\end{IEEEbiographynophoto}  
\vspace{-1cm}

\begin{IEEEbiographynophoto}{YUXIANG ZHANG (zhangyx@bupt.edu.cn)}is currently a associate researcher in Beijing University of Posts and Telecommunications.
\end{IEEEbiographynophoto}  
\vspace{-1cm}

\begin{IEEEbiographynophoto}{JIALIN WANG (wangjialinbupt@bupt.edu.cn)}is currently pursuing the
Ph.D. degree with Beijing University of Posts and Telecommunications.
\end{IEEEbiographynophoto}  
\vspace{-1cm}

\begin{IEEEbiographynophoto}{TAO JIANG (jiangtao@chinamobile.com)}is currently a member of China Mobile Research Institute.
\end{IEEEbiographynophoto}  
\vspace{-1cm}

\begin{IEEEbiographynophoto}{LIANG XIA (xialiang@chinamobile.com)}is currently a senior member of technical staff in the Future Mobile Technology Laboratory.
\end{IEEEbiographynophoto}  
\vspace{-1cm}

\begin{IEEEbiographynophoto}{PING ZHANG (pzhang@bupt.edu.cn)}is currently a professor with Beijing University of Posts and Telecommunications. 
\end{IEEEbiographynophoto}

\vfill

\end{document}